\documentclass[aps,pre,preprint,nofootinbib,eqsecnum,tightenlines,
showpacs,amsmath,titlepage]{revtex4}
\usepackage{graphicx}
\begin{document}

\vspace*{2cm}

\title{Casimir force for electrolytes}

\author{J. S. H{\o}ye}\email{johan.hoye@ntnu.no}

\affiliation{Department of Physics, Norwegian University of Science and
Technology, N-7491 Trondheim, Norway}

\date{\today}

\begin{abstract}

The Casimir force between a pair of parallell plates filled with ionic particles is considered. We use a statistical mechanical approach and consider the classical high temperature limit. In this limit the ideal metal result with no transverse electric (TE) zero frequency mode is recovered. This result has also been obtained by Jancovici and \v{S}amaj earlier. Our derivation differs mainly from the latter in the way the Casimir force is evaluated from the correlation function. By our approach the result is easily extended to electrolytes more generally. Also we show that when the plates are at contact the Casimir force is in accordance with the bulk pressure as follows from the virial theorem of classical statistical mechanics.

\end{abstract}


\maketitle

\section{Introduction}
\label{intro}

It is a pleasure to contribute this work to a festshrift volume for Professor Iver Brevik. We have had an extensive collaboration through many years on problems connected to the Casimir effect. In our works we have fruitfully utilized methods from different fields of research. In particular we have explored the statistical mechnanical aspects of the Casimir problem. The present contribution is a work that continues in the statistical mechanical direction.

A pair of metallic or dielectric plates attract each other. This is the well known Casimir effect, and it is commonly regarded to be due to fluctuations of the quantum electrodynamic field in vacuum. However, H{\o}ye and Brevik considered this in a different way by regarding the problem as a statistical mechanical one of interacting fluctuating dipole moments of polarizable particles. In this way the Casimir force between a pair of polarizable point particles was recovered \cite{brevikhoye88}. To do so the path integral formulation of quantized particle systems was utilized \cite{feynman53}. Before that this method was fruitfully utilized for a polarizable fluid \cite{hoyestell81}. With this approach the role of the electromagnetic field is to mediate the pair interaction between polarizable particles. Later this type of evaluation was generalized to a pair of parallell plates, and the well known Lifshitz result was recovered \cite{hoyebrevik98}. Similar evaluations were performed for other situations \cite{hoyebrevik00, hoyebrevik01}.

The statistical mechanical approach opens new perspectives for evaluations of the Casimir force. Instead of focusing upon the quantization of the electromagnetic field itself one can regard the problem as one of polarizable particles interacting via the electromagnetic field. It is found that these two viewpoints are equivalent \cite{brevikhoye88, hoyebrevik98, hoyebrevik01, hoyebrevikaarsethmilton03}.

Metals are materials that have electrons that can be regarded as free. When deriving the Lifshitz formula they are regarded as dielectric media that have infinite dielectric constant for zero frequency. Jancovici and \v{S}amaj realized that it should be possible to evaluate the Casimir force for metals by regarding an electron plasma. Thus they considered parallell plates filled with charged particles at low density in a neutralizing background \cite{jancovicisamaj04, jancovicisamaj05, buenzlimartin05}. Further they considered the classical case, i.e.~the high temperature limit. In this situation the Debye-H\"uckel theory of electrolytes is fully applicable. Then they use the Ornstein-Zernike equation (OZ) equation, and utilize its equivalence with the differential equation for the screened Coulomb potential to obtain the pair correlation function. This function is used to obtain the local ionic density at the surfaces of the plates. The difference between local and bulk densities is attributed to the Casimir force in accordance with the ideal gas law. The result obtained coincides with a result for ideal metals in the high temperature limit. The latter has been a dispute of controversy \cite{bordag00}. The ionic plasma result coincides with the one where there is no transverse electric mode at zero frequency. This is also in accordance with Maxwell's equations of electromagnetism.

The ionic plasma has also been extended to the quantum mechanical case by use of the path integral formalism from a statistical mechanical viewpoint, and it has been shown that magnetic interactions do not contribute in the classical high temperature limit \cite{buenzlimartin08}.

In the present work we reconsider the ionic plasma in the classical limit. We arrive at the the same pair correlation function as in Ref.~\cite{jancovicisamaj04}. But we use a different approach to obtain the Casimir force. As we see it, our method better utilizes the methods of classical statistical mechanics especially for possible further developments. Thus we use the correlation function to directly evaluate the average force between pairs of particles in the two plates and then integrate to obtain the total force. This is the method used in Refs.~\cite{brevikhoye88, hoyebrevik98}. In this way the result of Ref.~\cite{jancovicisamaj04} is recovered. A noteable feature of this comparison is that it demonstrates that the modification of the density profile at the surface is a perturbing effect that can be neglected to leading order by our approach.

With our approach the evaluations are extended in a straightforward way to electrolytes of more arbitrary density. To do so known properties of the direct correlation function is utilized. The main change with this extension is that the large distance inverse shielding length is modified while the Casimir force remains unchanged for large separations.

An additional result of our approach is that it is shown that when the plates are at contact the Casimir pressure more generally is nothing but the contribution to the bulk pressure (with opposite sign) that follows from the virial theorem of classical statistical mechnanics.

\section{General expressions}
\label{sec2}

Consider a pair of harmonic oscillators with static polarizability $\alpha$. They interact via a potential $\psi s_1 s_2$ where $s_1$ and $s_2$ are fluctuating polarizations. This interaction creates a shift in the free energy of the system. This is easily evaluated to be \cite{brevikhoye88}
\begin{equation}
-\beta F=-\frac{1}{2}\ln[1-(\alpha\psi)^2]=\frac{1}{2}\sum\limits_{n=1}^\infty\frac{1}{n}(\alpha\psi)^{2n}
\label{1}
\end{equation}
with $\beta=1/(k_B T)$ where $T$ is temperature and $k_B$ is Boltzmanns constant. The last sum is the expansion performed in Ref.~\cite{hoyebrevik98} where the two particles were replaced with two plane parallell plates. In the latter case the terms can be interpreted as the sum of graph contributions due to the mutual interaction $\psi$. The $\alpha$ will represent correlations within each plate separately while each $\psi$ gives a link between the plates while $2n$ is the symmetry factor of the graphs that form closed rings. With plates the endpoints of each link $\psi$ should be integrated over the plates. 
In the quantum mechanical case there is also a sum over Matzubara frequencies upon which $\alpha$ and $\psi$ may depend.

The parallell plates are separated by a distance $a$. Due to the interaction there will be an attractive force $K$ between the plates. This force is found from \cite{hoyebrevik98}
\begin{equation}
K=-\frac{\partial F}{\partial a}= \frac{1}{\beta}\frac{\alpha\psi\alpha}{1-(\alpha\psi)^2}\frac{\partial \psi}{\partial a}.
\label{2}
\end{equation}
The fraction in the middle of this expression represents the graph expansion of the pair correlation function with the endpoints in separate plates. These graphs form chains where each $\psi$ forms a link between the plates. Thus we can write
\begin{equation}
K=\rho^2\int h({\bf r}_2, {\bf r}_1)\,\psi_z'({\bf r}_2 - {\bf r}_1)\,d{\bf r}_1 d{\bf r}_2
\label{3}
\end{equation}
where $\rho$ is number density, $h({\bf r}_2, {\bf r}_1)$ is the pair correlation function, and $\psi_z'({\bf r}_2 - {\bf r}_1)=\partial\psi/\partial a$ with the $z$-direction normal to the plates. For polarizable particles integral (\ref{3}) will also contain integrations with respect to polarizations \cite{hoyebrevik98}.

For infinite plates integral (\ref{3}) diverges, so as usual we will consider the force $f$ per unit area which then will be
\begin{equation}
f=\frac{\rho^2}{(2\pi)^2}\int\limits_{z_1<0,z_2>0}\hat{h}(k_\perp,z_2,z_1)\hat{\psi}_z'(k_\perp,z_2-z_1)\,dk_x dk_y dz_1 dz_2
\label{4}
\end{equation}
where the hat denotes Fourier transform with respect to the $x$- and $y$-coordinates. (Here we have used $\int fg\,dxdy=\int\hat{f}\hat{g}\,dk_x dk_y/(2\pi)^2$ and translational symmetry along the xy-plane.)

Now we can introduce
\begin{equation}
q^2=k_\perp ^2= k_x^2+k_y^2,\quad \mbox{with }\quad dk_x dk_y=2\pi q\,dq.
\label{5}
\end{equation}
Further with $z_2=u_2+a$ and $z_1=-u_1$
we then get
\begin{equation}
f=\frac{\rho^2}{2\pi}\int\limits_{u_1,u_2>0}\hat{h}(q,z_2,z_1)\hat{\psi}_z'(q,z_2-z_1)q\,dq dz_1 dz_2
\label{6}
\end{equation}

An interesting feature of result (\ref{6}) or (\ref{4}) is that it is fully consistent with the virial theorem in statistical mechanics. This means that when the plates are at contact for $a=0$ the Casimir force equals the contribution to the pressure from the virial integral with pair interaction $\psi$. With $a=0$ translational symmetry is also present in the $z$-direction, so we have
\begin{equation}
f=\rho^2\int\limits_{z_1<0,z_2>0} h(|{\bf r}_2-{\bf r}_1|) \psi_z'(|{\bf r}_2-{\bf r}_1|)\,dxdy dz_1 dz_2.
\label{7}
\end{equation}
With new variable $z=z_2-z_1$ one can first integrate with respect to $z_2$ which then will be confined to the region $0\leq z_2 \leq z$. Thus with $\int_0^z\,dz_2=z$ we obtain (${\bf r}={\bf r}_2- {\bf r}_1$)
\begin{equation}
f=\rho^2\int\limits_{z>0} h(r) \psi_z'(r)\,d{\bf r}=\frac{\rho^2}{6}\int h(r){\bf r}\nabla\psi(r)\,d{\bf r}
\label{8}
\end{equation}
where symmetry with respect to the x-, y-, and z-directions and with respect to positive and negative $z$ is used. (It may be noted that the above is correct if the average of $\psi$ is zero. Otherwise the pair distribution function $1+h$ should be used. But for neutral plates as for dielectric plates with dipolar interaction this average will be zero.)

\section{Pair correlation function}
\label{sec3}

To obtain the correlation function we use the Ornstein-Zernike  (OZ) equation
\begin{equation}
h({\bf r}_2,{\bf r}_1)=c({\bf r}_2,{\bf r}_1)+\int c({\bf r}_2,{\bf r}')\rho({\bf r}') h({\bf r}',{\bf r}_1)\,d{\bf r}'
\label{9}
\end{equation}
which here has been extended to non-homogeneous fluids. The $c({\bf r})$ is the direct correlation function. For week long-range forces \cite{hemmer64} or to leading order the $c({\bf r})$ is related to the interaction in a simple way
\begin{equation}
c({\bf r}_2,{\bf r}_1)=-\beta\psi.
\label{9a}
\end{equation}
For plate separations beyond interparticle distances the $\psi$ will be small anyway. For a plasma at low density we can write for all $r$
\begin{equation}
c({\bf r}_2,{\bf r}_1)=c(r)=-\beta\frac{q_c^2}{r}, \quad ({\bf r}={\bf r}_2-{\bf r}_1)
\label{10}
\end{equation}
where $q_c$ is the ionic charge assuming one component for simplicity. (Here Gaussian units are used.) To keep the system neutral a uniform background is assumed. As noted in Ref.~\cite{jancovicisamaj04} the OZ-equation is now equivalent to Maxwells equation of electrostatics. The similar situation was utilized in Ref.~\cite{hoyebrevik98} for dipolar interactions. 

Since $\psi$ is the electrostatic potential from a charge one has
\begin{equation}
\nabla^2 c(r)=4\pi\beta q_c^2\delta({\bf r}).
\label{11}
\end{equation}
With this Eq.~(\ref{9}) can be rewritten as
\begin{equation}
\nabla^2 \Phi-4\pi\beta q_c^2 \rho({\bf r})\Phi=-4\pi\delta({\bf r}-{\bf r}_0), \quad  h({\bf r},{\bf r}_0)=-\beta q_c^2 \Phi\label{12},
\end{equation}
where ${\bf r}_2$ and ${\bf r}_1$ have been replaced by ${\bf r}$ and ${\bf r}_0$ respectively. In the present case with parallell plates the number density is
\begin{equation}
\rho({\bf r})=\left\{
\begin{array}{ll}
\rho,\quad & z<0\\
0,  \quad  & 0<z<a\\
\rho,\quad & a<z
\end{array}
\right.
\label{13}
\end{equation}
with equal densities $\rho={\rm const.}$ on both plates. By Fourier transform in the $x$- and $y$-directions Eq.~(\ref{12}) becomes
\begin{equation}
\left(\frac{\partial^2}{\partial z^2}-k_\perp^2-\kappa_z^2\right)\hat\Phi=-4\pi\delta(z-z_0)
\label{14}
\end{equation}
where the hat denotes Fourier transform and with $\kappa^2=4\pi\beta q_ c^2 \rho$

\begin{equation}
\kappa_z^2=\kappa^2\left\{
\begin{array}{ll}
1,\quad & z<0\\
0,  \quad  & 0<z<a\\
1,\quad & a<z. 
\end{array}
\right.
\label{15}
\end{equation}
The $\kappa$ is the inverse Debye-H{\" u}ckel shielding length in the media. Solution of Eq.~(\ref{14}) can be written in the form
\begin{equation}
\hat\Phi=2\pi e^{q_\kappa z_0}\left\{
\begin{array}{ll}
\frac{1}{q_\kappa}e^{-q_\kappa z}+Be^{q_\kappa z},\quad & z_0<z<0\\
Ce^{-qz}+C_1 e^{qz},  \quad  & 0<z<a\\
De^{-q_\kappa z},\quad & a<z
\end{array}
\right.
\label{16}
\end{equation}
where $q=k_\perp$, $q_\kappa=\sqrt{k_\perp^2+\kappa^2}$. (For $z<z_0$ the solution is the first line
of Eq.~(\ref{16}) where the resulting exponent of first exponential has changed sign.)

With continuous $\hat\Phi$ and $\partial\hat\Phi/\partial z$ as conditions, one finds for the coefficient of interest
\begin{equation}
D=\frac{4qe^{(q_\kappa -q)a}}{(q_\kappa+q)^2(1-Ae^{-2qa})}, \quad A=\left(\frac{q_\kappa-q}{q_\kappa+q}\right)^2=\frac{\kappa^4}{(q_k+q)^4}.
\label{17}
\end{equation}
With this the pair correlation function for $z_0<0$ and $z>a$ is
\begin{equation}
\hat h(k_\perp,z,z_0)=-2\pi\beta q_c^2 De^{-q_\kappa(z-z_0)}.
\label{18}
\end{equation}
%

\section{Casimir force}
\label{sec4}

Besides $\hat h$ the $\hat\psi_z'$ is needed to obtain the Casimir force $f$. In accordance with Eq.~(\ref{10}) the ionic pair interaction is $\psi=q_c^2/r$. Its full Fourier transform is $\tilde\psi=4\pi q_c^2/k^2$ which is consistent with Eq.~(\ref{11}). With $k^2=k_\perp^2+k_z^2$ this can be transformed backwards to obtain ($q=k_ \perp$)
\begin{equation}
\hat \psi(k_\perp,z-z_0)=2\pi q_c^2 \frac{e^{-q(z-z_0)}}{q}.
\label{19}
\end{equation}
This is consistent with solution (\ref{16}) for $\Phi$. The derivative of (\ref{19}) with respect to $z$ is now together with expression (\ref{18}) inserted in Eq.~(\ref{6}) to first obtain ($z-z_0\rightarrow z_2-z_1=u_1+u_2+a$)
\begin{eqnarray}
f&=&\frac{\rho^2}{2\pi}\int\limits_0^\infty(-2\pi\beta q_c^2)D(2\pi q_c^2)\int\limits_0^\infty\int\limits_0^\infty e^{-(q_\kappa+q)(u_1+u_2+a)}\,du_1 du_2 \,q\,dq
\nonumber\\
&=&-\frac{\kappa^4}{8\pi\beta}\int\limits_0^\infty
\frac{De^{-(q_\kappa+q)a}}{(q_\kappa+q)^2}q\,dq=
-\frac{1}{2\pi\beta}\int\limits_0^\infty\frac{Ae^{-2qa}}{1-Ae^{-2qa}}q^2\,dq.
\label{20}
\end{eqnarray}
First one can note that this result is precisely result (3.44) in Ref.~\cite{jancovicisamaj04}. This is seen by some rearrangement of the latter result with the substitutions $\kappa_0\rightarrow\kappa$, $k\rightarrow q/\kappa$, and $d\rightarrow a$ for dimensionality $\nu=3$.

Expression (\ref{17})  and result (\ref{20} may be simplified further with new variable of integration
\begin{equation}
q=\kappa \sinh t,  \quad dq=\kappa \cosh{t}\,dt.
\nonumber
\end{equation}
With this we have $q_\kappa+q=\kappa(\cosh t+\sinh t)=\kappa e^t$, $q_\kappa-q=\kappa e^{-t}$, and $A=e^{-4t}$ by which the Casimir force becomes
\begin{equation}
f=\frac{\kappa^3}{2\pi\beta}\int\limits_0^\infty\frac{e^{-g(t)}}{1-e^{-g(t)}}\sinh^2 t\cosh t\, dt.
\label{21}
\end{equation}
where $g(t)=4t+2\kappa a\sinh t$.

For large separation $a$ only small values of $t$ will contribute, and one can put
\begin{equation}
g(t)=(2\kappa a+4)t \quad {\rm and} \quad \sinh^2 t\cosh t=t^2.
\nonumber
\end{equation}
With this and expansion of the denominator the force becomes
\begin{equation}
f=\frac{\kappa^3}{2\pi\beta}\frac{2\zeta(3)}{(2\kappa a+4)^3}=\frac{k_B T\zeta(3)}{8\pi a^3(1+2/(\kappa a))^3}=\frac{k_B T\zeta(3)}{8\pi a^3}\left(1-\frac{6}{\kappa a}+\cdots\right).
\label{22}
\end{equation}
The $\zeta(3)$ is the Riemann $\zeta$-function, $\zeta(p)=\sum_{n=1}^\infty 1/n^p$.

As noted earlier \cite{jancovicisamaj04,jancovicisamaj05,buenzlimartin05} this is the ideal metal result for high temperatures when the transverse electric mode is absent. Also one sees that for large $a$ the effective separation between the plates is increasesd by twice the Debye shielding length, i.~e.~$a\rightarrow a+2/\kappa$. Thus for semiconductors the influence of free ions vanishes due to the increase of effective separation for decreasing ionic density. The small conductivity of semiconductors has been an issue of some controversy \cite{ellingsenbrevikhoyemilton08}. It has been argued that small concentrations of free ions in semiconductors should be neglected \cite{geyer05}. However, result (\ref{19}) suggests that lack of influence for small ionic concentration is due to increased effective separation for vanishing $\kappa$.

When the plates are in contact, $a=0$, the integral (\ref{18}) can be evaluated exlicitly. With $1-e^{-4t}=4e^{-2t}\sinh t\cosh t$ one finds
\begin{equation}
f=\frac{\kappa^3}{8\pi\beta}\int\limits_0^\infty e^{-2t}\sinh t\,dt=\frac{\kappa^3}{24\pi\beta}.
\label{23}
\end{equation}
For an ionic system at low density this is precisely the contribution to the pressure (with opposite sign) from the ionic interaction (beyond the ideal gas pressure) in accordance with the virial integral (\ref{8}).

\section{Electrolytes in general}
\label{sec5}

For higher densities and lower temperatures the direct correlation function $c$ given by Eq.~(\ref{10}) will be modified. However, the crucial point is that for large $r\rightarrow \infty$ this expression is still valid while for small $r$ there will be changes. On the scale of plate separation this change will be a term that can be regarded as a $\delta$-function in ${\bf r}$-space such that 
\begin{equation}
c({\bf r}_2,{\bf r}_1)=c_0(r)+\tau\delta({\bf r}_2-{\bf r}_1)
\label{24}
\end{equation}
where $c_0(r)=-\beta q_c^2/r$ and $\tau$ is a constant that will depend upon the local density.
When the local density varies the OZ-equation (\ref{9}) can be regarded as a matrix equation. Multiplying it from both left and right with $\rho$ and adding $\rho$ on both sides of it the equation after some rearrangement becomes
\begin{equation}
(1-\rho c)\rho(1+h\rho)=\rho.
\label{25}
\end{equation}
Insertion of expression (\ref{24}) then yields
\begin{equation}
(1-\rho\tau-\rho c_0)\rho(1+h\rho)=\rho.
\nonumber
\end{equation}
\begin{equation}
\rho(1+h\rho)=\frac{\rho_e}{1-\rho_e c_0},\quad\rho_e=\frac{\rho}{1-\rho\tau}.
\label{26}
\end{equation}
Thus the only change in the resulting pair corrrelation function $\rho h\rho$ is that $\rho$ is replaced by an effective density $\rho_e$ on the right hand side. In this way only the inverse shielding length is affected by which we get $\kappa^2=4\pi\beta\rho_e q_c^2$. But for large plate separations the Casimir force (\ref{22}) does not depend upon $\kappa$ by which the ideal metal result is generally valid for large separations for any electrolyte.

\section{Summary}
\label{sec6}

The Casimir force between a pair of parallell plates filled with ionic particles has been evaluated in the classical high temperature limit. To do so methods of classical statistical mechanics have been used. The pair correlation function is evaluated from which the average force between pairs of particles in different plates is found. When the plates are at contact the magnitude of the force equals the contribution to the pressure from the virial theorem. This latter result makes the force consistent with bulk pressure. The force found is the same as the one found earlier in Ref.~\cite{jancovicisamaj04} for charged particles at low density. There the force was evaluated on basis of the difference between surface and bulk densities. By the present approach it thus follows that this difference in densities can be neglected to leading order.


\end{document}